\documentclass[twocolumn,showpacs,tighten,floats,amsmath,amssymb,prb]{revtex4}
\usepackage{graphicx}    % Include figure files
\usepackage{epsfig}      % Include figure files
\usepackage{dcolumn}     % Align table columns on decimal point
\usepackage{bm}          % Bold math
\usepackage{subfigure}
\newcommand{\LAST}{Ag$_{x}$Pb$_{m}$SbTe$_{m+2}$}
\newcommand{\ptps}{(PbTe)$_{1-x}$(PbS)$_{x}$}

\newcommand{\ptpstwo}{PbTe$_{0.84}$S$_{0.16}$}

\newcommand{\ptpsfour}{PbTe$_{0.5}$S$_{0.5}$}

\begin{document}
%%%%%%%%%%%%%%%%%%%%%%%%%%%%%%%%%%%%%%%%%%%%%%%%%%%%%%%%%%%%%%%%%%%%%%%%%%%
\title{Phase separation and nanostructuring in the thermoelectric material PbTe$_{1-x}$S$_x$}
\author {H. Lin, E. S. Bo\v zin and S. J. L. Billinge}
\email{billinge@pa.msu.edu}
\homepage{http://nirt.pa.msu.edu/}
\affiliation{Department of Physics and Astronomy, Michigan State
University, East Lansing, MI 48824}
\author {J. Androulakis, C. H. Lin and M. G. Kanatzidis}
\affiliation{Department of Chemistry, Michigan State University,
East Lansing, MI 48824}
\date{\today}
\begin{abstract}
   The average and local structures of the \ptps\ system of thermoelectric
materials has been studied using the Rietveld and atomic pair
distribution function (PDF) methods.  Samples with $0.25\le x$ are
macroscopically phase separated. Phase separation was suppressed in
a quenched $x=0.5$ sample which, nonetheless, exhibited a partial
spinodal decomposition.  The promising thermoelectric material with
$x=0.16$ showed intermediate behavior.  Combining TEM and bulk
scattering data suggests that the sample is a mixture of PbTe rich
material and a partially spinodally decomposed phase similar to the
quenched 50\% sample.  This results in a nano-meter scale
inhomogeneous material that accounts for its very low thermal
conductivity.
\end{abstract}
\maketitle
%
%
%%%%%%%%%%%%%%%%%%%%%%%%%%%%%%%%%%%%%%%%%%%%%%%%%%%%%%%%%%%%%%%%%%%%%%%%%%%
% introduction
%%%%%%%%%%%%%%%%%%%%%%%%%%%%%%%%%%%%%%%%%%%%%%%%%%%%%%%%%%%%%%%%%%%%%%%%%%%
%
%
\section{Introduction}

Thermoelectric materials are the subject of intense research because
of their potential for efficient power generation and cooling. The
efficiency of the thermoelectric material is measured by the figure
of merit, $ZT$, defined by several interdependent physical
parameters.\cite{ZTnote} It is difficult to get a high ZT material
due to the competing requirements for optimizing the interdependent
parameters. Many efforts have focused on reducing the thermal
conductivity $\kappa$, without sacrificing electrical conductivity,
$\sigma$. $\kappa$ is the sum of the lattice thermal conductivity
$\kappa_{lat}$ and the electronic thermal conductivity
$\kappa_{ele}$. Theoretical and experimental studies suggest that
materials that show nano-phase separation appear to be promising in
achieving high
performance.\cite{harma;s02,harma;jem05,hsu;s04,kimw;prl06}

The material with composition PbTe$_{0.84}$S$_{0.16}$ shows a very
low room temperature lattice thermal conductivity of 0.4~W/m~K and a
$ZT$ value significantly higher than that of PbTe and
PbS.\cite{andro;unpub06} The thermal conductivity is only 28\% of
that observed in the PbTe system, which is remarkable given that the
two are isostructural and PbTe$_{0.84}$S$_{0.16}$ has only 16 At.~\%
of S substituted on the Te site. Understanding the origin of this
remarkable reduction in $\kappa$ for a small doping change should
give important insights into the thermoelectric problem.

Early studies on the \ptps\ system showed that phase separation
occurs at low temperature over almost the whole composition
range.\cite{darro;tmsaime66,leute;zpc85} A miscibility gap exists
over a wide range of composition and extends almost up to the
melting point of the alloy.  There are no apparent intermediate
compounds and the phase separation occurs into phases which are
almost pure PbTe and PbS over the whole alloy range. Theoretical
work\cite{leute;zpc85} supports such a picture and the calculated
phase diagram using a thermodynamic model agreed with the previous
experimental data. Earlier
work\cite{yamam;srtu56,sinde;dansssr59,malev;dansssr63} suggested a
 smaller range for the miscibility gap in the phase diagram and
this discrepancy was attributed to the subtle difference in chemical
processing\cite{darro;tmsaime66} and quenching rate.
It is apparent from the high resolution transmission electron
microscopy (HRTEM) images is that phase separation occurs on several
different length-scales in \ptpstwo\ and that naturally forming
striped nanostructures due to spinodal decomposition are evident in
portions of the sample. Here we investigate this question further
using bulk diffraction probes of the average and local atomic
structure. We address two questions. First, can we confirm that the
nano-scale phase separation is a bulk property and can we
characterize the average chemical composition and structure of the
spinodal domain? We have also extended the study to other
compositions in the phase diagram to see how these effects evolve
with changing composition.

The atomic pair distribution function (PDF) analysis of x-ray
diffraction data is a useful method for studying nano-phase
separated samples.\cite{billi;cc04,egami;b;utbp03}  In the PDF
approach both Bragg and diffuse scattering are analyzed and it
yields the bulk average local atomic structure. Recently it was
successfully used to study the thermoelectric material, \LAST, where
silver and antimony rich nano-scale clusters were found to be
coherently embedded in the PbTe matrix as a bulk
property.\cite{lin;prb05}

We have used both PDF and Rietveld methods to study the \ptps\
system. We find phase separation occurring over the whole
composition range. Refinements from both Rietveld and PDF methods
show that the $x=0.25$, 0.5, and 0.75 samples are macroscopically
separated into phases that are almost pure PbS and PbTe.  This does
not happen in the important 16\% PbS doped sample. However, taking
all the evidence together we suggest that the 16\% sample is a
nanoscale mixture of a PbTe rich phase with a partially spinodally
decomposed phase of nominally 50\% composition. Such a phase was
stabilized and observed in a quenched $x=0.5$ sample in this study.
This offers the opportunity in the future for engineering nano- and
micro-structures with favorable thermoelectric properties by
controlling the thermal history in these materials.
%
%
%%%%%%%%%%%%%%%%%%%%%%%%%%%%%%%%%%%%%%%%%%%%%%%%%%%%%%%%%%%%%%%%%%%%%%%%%%%%
% Experimental methods
%%%%%%%%%%%%%%%%%%%%%%%%%%%%%%%%%%%%%%%%%%%%%%%%%%%%%%%%%%%%%%%%%%%%%%%%%%%%
%
%
\section{Experimental methods}
Powder samples in the \ptps\ series were made with different
compositions: $x=0$, 0.16, 0.25, 0.50, 0.75 and 1. The samples were
produced by mixing appropriate ratios of high purity elemental
starting materials with a small molar percentage of PbI$_{2}$, an
$n$-type dopant. The initial loads were sealed in fused silica tubes
under vacuum and fired at 1273~K for 6~h, followed by rapid cooling
to 773~K and held there over a period of 72~h.  One $x=0.5$ sample
was also quenched rapidly to room-temperature. More details of
sample synthesis can be found elsewhere.\cite{andro;unpub06}

Finely powdered samples were packed in flat plates with a thickness
of 1.0~mm sealed between kapton tape windows. X-ray powder
diffraction data were collected using the rapid acquisition PDF
(RA-PDF) method,\cite{chupa;jac03} which benefits from very high
energy x-rays and a two-dimensional detector.  The experiments were
conducted using synchrotron x-rays with an energy of 86.727~keV
($\lambda=0.14296$~\AA) at the 6-ID-D beam line at the Advanced
Photon Source (APS) at Argonne National Laboratory. The data were
collected using a circular image plate camera (Mar345) 345~mm in
diameter. The camera was mounted orthogonally to the beam path with
a sample-to-detector distance of 210.41~mm.

\begin{figure}[tb]
\centering
\includegraphics[width=0.45\textwidth,angle=0]{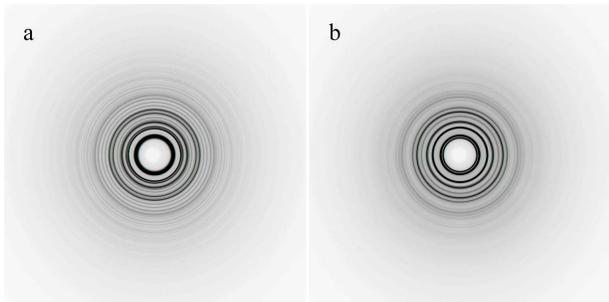}
\caption{Raw x-ray powder diffraction data from the 2D detector for
the $x=0.50$ \ptps ~sample. Data from the (a) unquenched and (b)
quenched samples are shown for comparison.  The 1-D integrated
powder diffraction patterns obtained from these data are shown in
Fig.~\ref{fig;all1}(a) and on an expanded scale in
Fig.~\ref{fig;FqAllLowQ}. The white circle in the center of each 2D
diffractogram represents a shadow from the beam-stop.}
\label{fig;two-images}
\end{figure}

In order to avoid saturation of the detector, each room temperature
measurement was carried out in multiple exposures. Each exposure
lasted 5 seconds, and each sample was exposed five times to improve
the counting statistics. Two representative 2D diffraction images
for unquenched and quenched \ptpsfour\ samples are shown in
Fig.~\ref{fig;two-images}(a) and (b), respectively.
%
%%%%%%%%%%%%%%%%%%%%%%%%%%%%%%%%%%%%%%%%%%%%%%%%%%%%%%%%%%%%%%%%%%%
% Start Figure 1
%%%%%%%%%%%%%%%%%%%%%%%%%%%%%%%%%%%%%%%%%%%%%%%%%%%%%%%%%%%%%%%%%%%
%%%%%%%%%%%%%%%%%%%%%%%%%%%%%%%%%%%%%%%%%%%%%%%%%%%%%%%%%%%%%%%%%%%
% End Figure 1
%%%%%%%%%%%%%%%%%%%%%%%%%%%%%%%%%%%%%%%%%%%%%%%%%%%%%%%%%%%%%%%%%%%
%
The excellent powder statistics, giving uniform rings, are evident.
All the samples yielded similar quality images.  The 2D Data sets
from each sample were combined and integrated using the program
FIT2D\cite{hamme;esrf98} before further processing.

\begin{figure}[tb]
\centering
\includegraphics[width=0.35\textwidth,angle=270]{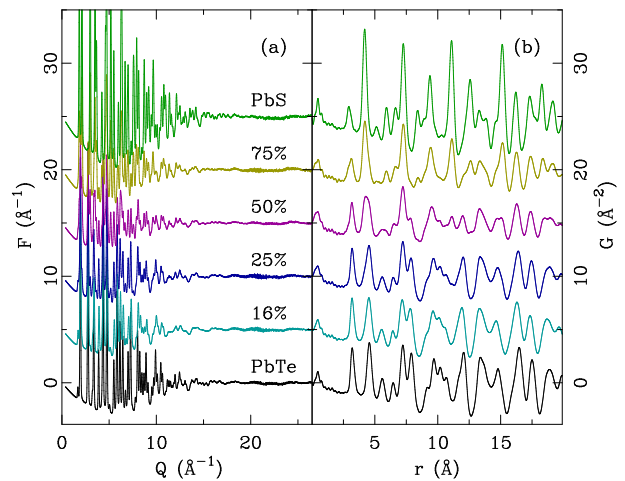}
\caption{Experimental (a) $F(Q)$ and (b) $G(r)$ for all unquenched samples.
In the Fourier transform, $Q_{max}$ was set to 26.0~\AA$^{-1}$.  The
data are offset for clarity.  The compositions of the \ptps ~samples are
indicated in panel (a). From top to bottom: $x=1.00$ (green), $x=0.75$
(yellow), $x=0.50$ (magenta), $x=0.25$ (blue), $x=0.16$ (cyan), and
$x=0.00$ (black).}
\label{fig;all1}
\end{figure}

Data from an empty container were also collected to subtract the
container scattering.  The corrected total scattering structure
function, $S(Q)$, was obtained using standard
corrections\cite{egami;b;utbp03,chupa;jac03} with the program
PDFgetX2.\cite{qiu;jac04i}  Finally, the PDF was obtained by Fourier
transformation of $S(Q)$ according to
$G(r)=\frac{2}{\pi}\int_{0}^{Q_{max}} Q [S(Q)-1] \sin (Qr)\> dQ$,
where $Q$ is the magnitude of the scattering vector. A
$Q_{max}=26.0$~\AA$^{-1}$ was used. Fig.~\ref{fig;all1}, shows
$F(Q)=Q (S(Q)-1)$ and $G(r)$ for all the samples.
%
%%%%%%%%%%%%%%%%%%%%%%%%%%%%%%%%%%%%%%%%%%%%%%%%%%%%%%%%%%%%%%%%%%%
% Start Figure 2
%%%%%%%%%%%%%%%%%%%%%%%%%%%%%%%%%%%%%%%%%%%%%%%%%%%%%%%%%%%%%%%%%%%
%%%%%%%%%%%%%%%%%%%%%%%%%%%%%%%%%%%%%%%%%%%%%%%%%%%%%%%%%%%%%%%%%%%
% End Figure 2
%%%%%%%%%%%%%%%%%%%%%%%%%%%%%%%%%%%%%%%%%%%%%%%%%%%%%%%%%%%%%%%%%%%
%
The good statistics and overall quality of the data are apparent in
Fig.~\ref{fig;all1}(a).  The low spurious ripples at low-$r$ in the
$G(r)$ functions are also testament to the quality of the
data.\cite{peter;jac03} Note that $G(r)$ has been plotted all the
way to $r=0$ in these plots, which is a stringent test of this.

%
%
%%%%%%%%%%%%%%%%%%%%%%%%%%%%%%%%%%%%%%%%%%%%%%%%%%%%%%%%%%%%%%%%%%%
% Modeling
%%%%%%%%%%%%%%%%%%%%%%%%%%%%%%%%%%%%%%%%%%%%%%%%%%%%%%%%%%%%%%%%%%%
%
%
\section{Modeling}
Both PDF (using the PDFfit2 program\cite{farro;jpcm06,proff;jac99})
and Rietveld\cite{rietv;jac69} (using the TOPAS
program\cite{coelh;unpub04}) refinements were carried out on the
system. The models used in the fits are described below.

One of the main outcomes of this study is to determine the phase
composition of the phase-separated sample as a function of
composition. When phase separation is long-ranged, Rietveld
refinement can be used to estimate the relative abundance of the
phase
components\cite{werne;jac79,hill;jac87,bish;jac88,conno;pd88,hill;pd91,young;b;trm93}.

Phase segregation can also be determined from the
PDF.\cite{lin;prb05,proff;zk02} In PDFfit2, each phase in a
multi-phase fit has its own scale-factor in the refinement. The
scale factor reflects both the relative phase-fraction of the phases
and the average scattering power of each phase, which depends on the
chemical compositions of each phase. The conversion from
scale-factor to atomic-fraction is done using the equations derived
in Ref.~\onlinecite{lin;prb05}.

For each sample, we explored different models. The structure is of
the rock-salt type, space group Fm-3m.  First we start from a
homogeneous (solid solution) model where the anions are assumed to
randomly distributed on the sites of the anionic sublattice. In this
model, S atoms substitute the Te site randomly without breaking the
symmetry. The only structural parameters refined are the lattice
constants and the atomic displacement factors.

The next model we tried was a simple two-phase model in which a
phase separation into a PbTe-rich and PbS-rich phase was assumed.
The phase diagram for this system shows a miscibility gap at low
temperature over a wide composition
range.\cite{darro;tmsaime66,leute;zpc85}  The two phases that
coexist have compositions rather close to the pure end-members and
there is limited solid solubility.  Based on this, and in an effort
to keep the modeling as simple as possible, we modeled the phase
separation as a mixture of pure PbTe and PbS; however, allowing the
lattice constants to vary as would be expected if the phases were
not the pure end-members. The parameters that were allowed to vary
in these fits were lattice constants, atomic displacement factors
and phase specific scale factors which reflect the relative
abundance of each phase. More complicated phase separated models
were also tried where the composition of the phases was varied as
described below.

%
%
%%%%%%%%%%%%%%%%%%%%%%%%%%%%%%%%%%%%%%%%%%%%%%%%%%%%%%%%%%%%%%%%%%%
% Results
%%%%%%%%%%%%%%%%%%%%%%%%%%%%%%%%%%%%%%%%%%%%%%%%%%%%%%%%%%%%%%%%%%%
%
%
\section{Results}
\begin{figure}[tb]
\centering
\includegraphics[width=0.35\textwidth,angle=270]{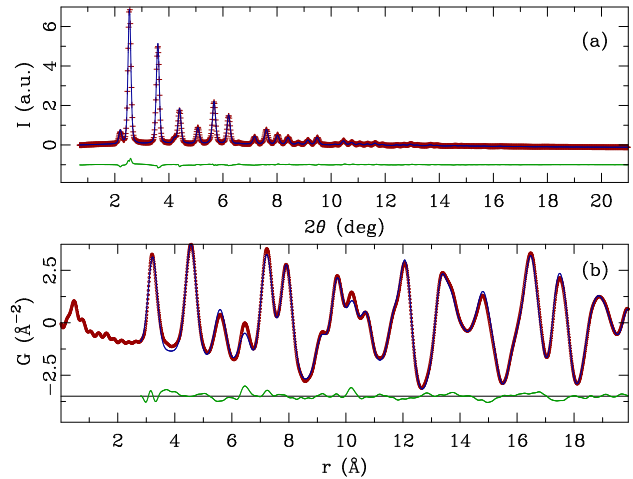}
\caption{Representative refinements of the PbTe data using (a)
Rietveld and (b) PDF approaches. Symbols represent data, and solid
lines are the model fits. The difference curves are offset for
clarity.} \label{fig;exampleFits}
\end{figure}

First we carried out PDF and Rietveld refinements on the undoped
end-members of the series, PbS and PbTe. The level of agreement of
Rietveld and PDFfit refinements can be seen in
Fig.~\ref{fig;exampleFits}
%
%%%%%%%%%%%%%%%%%%%%%%%%%%%%%%%%%%%%%%%%%%%%%%%%%%%%%%%%%%%%%%%%%%%
% Start Figure 3
%%%%%%%%%%%%%%%%%%%%%%%%%%%%%%%%%%%%%%%%%%%%%%%%%%%%%%%%%%%%%%%%%%%
%%%%%%%%%%%%%%%%%%%%%%%%%%%%%%%%%%%%%%%%%%%%%%%%%%%%%%%%%%%%%%%%%%%
% End Figure 3
%%%%%%%%%%%%%%%%%%%%%%%%%%%%%%%%%%%%%%%%%%%%%%%%%%%%%%%%%%%%%%%%%%%
%
and Table~\ref{tab;end-members-lattice}.
%
%%%%%%%%%%%%%%%%%%%%%%%%%%%%%%%%%%%%%%%%%%%%%%%%%%%%%%%%%%%%%%%%%%%
%  Start Table 1
%%%%%%%%%%%%%%%%%%%%%%%%%%%%%%%%%%%%%%%%%%%%%%%%%%%%%%%%%%%%%%%%%%%
\begin{table}[tbp]
\centering \caption{Refinement results from PbS and PbTe compared
with literature values.}
\begin{tabular}{|c|c|c|c|}
\hline &Literature\tablenote{Ref.~\onlinecite{noda;acc87}}&Rietveld&PDF\\
\hline$R_w$&-&0.03994&0.0852\\
$a_{PbTe}$ (\AA)&6.4541(9)&6.4776(3)&6.465(3)\\
$U_{Pb}$ (\AA$^2$)& 0.0204(3) &0.033(5)&0.032(4)\\
$U_{Te}$  (\AA$^2$)& 0.0141(2) &0.009(9)&0.014(4)\\
\hline $R_w$&-&0.04377&0.0820\\
$a_{PbS}$ (\AA)&5.9315(7)&5.9460(3)&5.940(3)\\
$U_{Pb}$ (\AA$^2$)&0.0163(3)  &0.023(3)&0.0185(5)\\
$U_{S}$ (\AA$^2$)& 0.0156(5) &0.018(4)&0.030(5)\\
\hline
\end{tabular}
\label{tab;end-members-lattice}
\end{table}
%%%%%%%%%%%%%%%%%%%%%%%%%%%%%%%%%%%%%%%%%%%%%%%%%%%%%%%%%%%%%%%%%%%
%  End Table 1
%%%%%%%%%%%%%%%%%%%%%%%%%%%%%%%%%%%%%%%%%%%%%%%%%%%%%%%%%%%%%%%%%%%
%
These fits give a baseline for the quality of the fits for materials
without disorder. The fits are acceptable and the refined parameters
are in reasonable agreement with literature values for PbTe, though
outside the estimated errors. The PDF and Rietveld refinements are
also only in semi-quantitative agreement. The parameter estimates
were made on the same data-sets but using different methods and
systematic errors are not accounted for in the error estimates. Even
in these nominally pure materials the refined atomic displacement
factors are rather large\cite{jeong;jpc99}, which is in agreement
with previous work,\cite{noda;acc87} though this behavior is not
really understood.

\begin{figure}[tb]
\centering
\includegraphics[width=0.425\textwidth,angle=270]{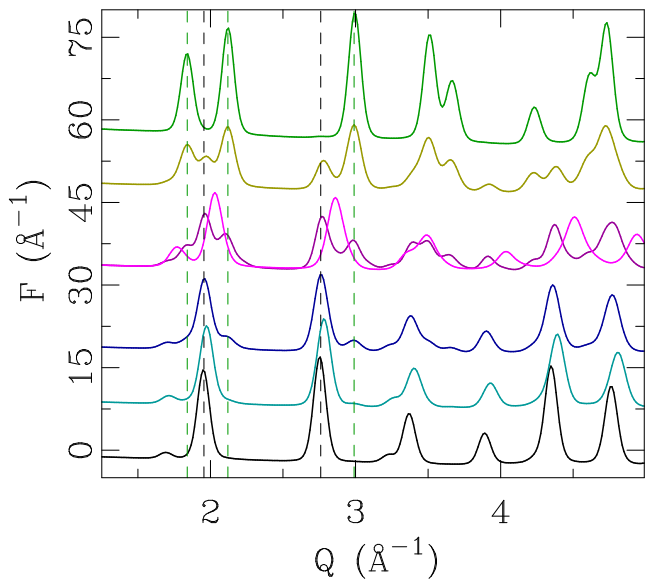}
\caption[]{The low-$Q$ diffraction patterns of all the \ptps\ samples
studied, where $F(Q)=Q(S(Q)-1)$. From top to bottom: $x=1.00$
(green), $x=0.75$ (yellow), $x=0.50$ (light and dark magenta),
$x=0.25$ (blue), $x=0.16$ (cyan), and $x=0.00$ (black). The data
corresponding to the quenched $x=0.50$ sample (light magenta) is
superimposed on top of that of the unquenched sample (dark magenta)
without being offset. The other data are offset for clarity.
Vertical dashed lines indicate positions of several characteristic
Bragg peaks in the end-member data to allow for easier comparison.}
\label{fig;FqAllLowQ}
\end{figure}

Now we consider the chemically mixed systems. The existence of phase
separation can be \emph{qualitatively} verified in our samples by
looking at the diffraction patterns in Fig.~\ref{fig;FqAllLowQ}.
%
%%%%%%%%%%%%%%%%%%%%%%%%%%%%%%%%%%%%%%%%%%%%%%%%%%%%%%%%%%%%%%%%%%%
% Start Figure 4
%%%%%%%%%%%%%%%%%%%%%%%%%%%%%%%%%%%%%%%%%%%%%%%%%%%%%%%%%%%%%%%%%%%
%%%%%%%%%%%%%%%%%%%%%%%%%%%%%%%%%%%%%%%%%%%%%%%%%%%%%%%%%%%%%%%%%%%
% End Figure 4
%%%%%%%%%%%%%%%%%%%%%%%%%%%%%%%%%%%%%%%%%%%%%%%%%%%%%%%%%%%%%%%%%%%
%
The top curve is PbS and the bottom curve is PbTe and the vertical
dashed lines are at the positions of the main Bragg-peaks of these
phases. Despite the low-resolution of the data, a characteristic of
the RAPDF measurement,\cite{chupa;jac03} for compositions $x=0.25$,
0.50 and 0.75 a coexistence of PbS and PbTe diffraction patterns is
clearly evident as the diffraction patterns are qualitatively
recognizable as a linear superposition of the end-member patterns.
Diffraction peaks appear at precisely the positions of the
end-member Bragg-peaks.  The same is true for the annealed $x=0.5$
sample (dark magenta).  On the other hand, the \emph{quenched}
$x=0.5$ sample has a diffraction pattern that resembles the PbTe
pattern but shifted significantly to higher scattering angles.  This
is what would be expected for a solid-solution, rather than phase
separated, sample suggesting that quenching the sample suppresses
phase separation.

 The situation is slightly less clear for the
$x=0.16$ sample which resembles closely the pure PbTe diffraction
pattern. The effects of phase separation would be difficult to see
in this case because of the small PbS component.   However, careful
inspection of the curve indicates that the main peaks are shifted to
higher scattering angles, in analogy with the quenched $x=0.5$
sample. Thus, this sample appears to be a solid-solution on the
macro-scale probed in a diffraction pattern.

\begin{figure}[tb]
\centering
\includegraphics[width=0.45\textwidth,angle=270]{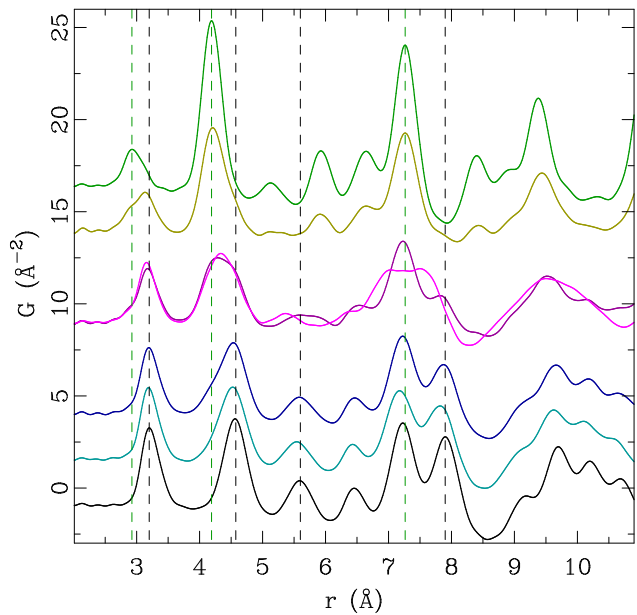}
\caption{Experimental PDFs for various \ptps\ samples on expanded
scale. The PDFs, from top to bottom correspond to $x=1.00$ (green),
$x=0.75$ (yellow), $x=0.50$ (magenta), quenched $x=0.50$ (bright
magenta), $x=0.25$ (blue), $x=0.16$ (cyan), and $x=0.00$ (black).
The data corresponding to the quenched $x=0.50$ sample (light
magenta) is superimposed on top of that of the unquenched sample
(dark magenta) without being offset. The other data are offset for
clarity. Vertical dashed lines indicate positions of a few selected
characteristic PDF features of the end-members for easier
comparison.} \label{fig;diff1}
\end{figure}

We would like to consider evidence in the \emph{local} structure for
phase separation.  The PDFs of the data in Fig.~\ref{fig;FqAllLowQ}
are shown in Fig.~\ref{fig;diff1} arranged in the same way and with
the same colors as in Fig.~\ref{fig;FqAllLowQ}.
%
%%%%%%%%%%%%%%%%%%%%%%%%%%%%%%%%%%%%%%%%%%%%%%%%%%%%%%%%%%%%%%%%%%%
% Start Figure 5
%%%%%%%%%%%%%%%%%%%%%%%%%%%%%%%%%%%%%%%%%%%%%%%%%%%%%%%%%%%%%%%%%%%
%%%%%%%%%%%%%%%%%%%%%%%%%%%%%%%%%%%%%%%%%%%%%%%%%%%%%%%%%%%%%%%%%%%
% End Figure 5
%%%%%%%%%%%%%%%%%%%%%%%%%%%%%%%%%%%%%%%%%%%%%%%%%%%%%%%%%%%%%%%%%%%
%
The samples that are macroscopically phase separated ($x=0.25,$ 0.5
(annealed) and 0.75) also show phase separation in the local
structure as expected, the curves having the qualitative appearance
of a mixture of the end-member PDFs.

The behavior of peaks in the PDF in solid-solutions has been
discussed previously.\cite{petko;prl99,jeong;prb01}  The nearest
neighbor peaks retain the character of the end-members, albeit with
a small strain relaxation.  However, peaks at higher-$r$, from the
second-neighbor onwards, appear broadened because of inhomogeneous
strain in the sample but are peaked at the average position expected
from the average structure for the solid solution. The $x=0.5$
(quenched) and $x=0.16$ samples follow this behavior even on the
1~nm length-scale suggesting that they are solid-solutions even on
the local scale.

\begin{figure}[tb]
\centering
\includegraphics[width=0.35\textwidth,angle=270]{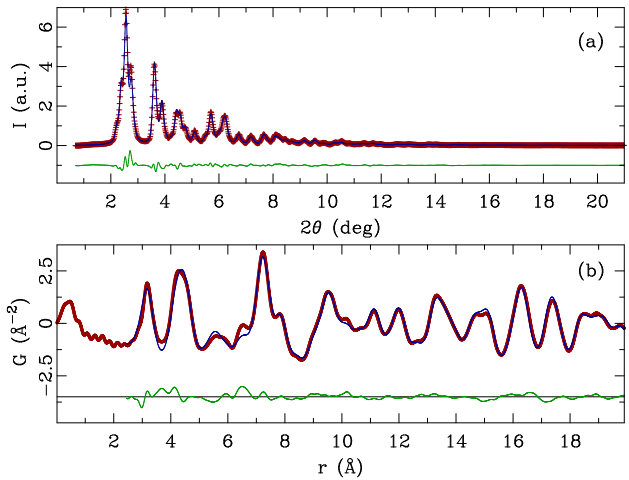}
\caption{Representative refinements of the $x=0.50$ sample data
using (a) Rietveld and (b) PDF approach. Symbols represent data,
and solid lines are the model fits. The difference curves are offset
for clarity.}
\label{fig;fiftyFits}
\end{figure}

To investigate the phase separation phenomenon more quantitatively,
we carried out two-phase refinements for the macroscopically phase
separated samples on both the diffraction data and the PDF.
Fig.~\ref{fig;fiftyFits} shows representative fits from the $x=0.50$
sample.
%%%\nb{x is the PbS composition.}
%
%%%%%%%%%%%%%%%%%%%%%%%%%%%%%%%%%%%%%%%%%%%%%%%%%%%%%%%%%%%%%%%%%%%
% Start Figure 6
%%%%%%%%%%%%%%%%%%%%%%%%%%%%%%%%%%%%%%%%%%%%%%%%%%%%%%%%%%%%%%%%%%%
%%%%%%%%%%%%%%%%%%%%%%%%%%%%%%%%%%%%%%%%%%%%%%%%%%%%%%%%%%%%%%%%%%%
% End Figure 6
%%%%%%%%%%%%%%%%%%%%%%%%%%%%%%%%%%%%%%%%%%%%%%%%%%%%%%%%%%%%%%%%%%%
%
The refined parameters are reproduced in
Table~\ref{tab;PDF-Rietveld-Nominal}. In the table the $n$ and
$n_{0}$ refer to the refined fraction of the sample in the PbTe
phase and the expected fraction based on the stoichiometry and
assuming phase separation into pure PbTe and PbS, respectively.
%
%%%%%%%%%%%%%%%%%%%%%%%%%%%%%%%%%%%%%%%%%%%%%%%%%%%%%%%%%%%%%%%%%%%
%  Start Table 2
%%%%%%%%%%%%%%%%%%%%%%%%%%%%%%%%%%%%%%%%%%%%%%%%%%%%%%%%%%%%%%%%%%%
\begin{table*}[tbp]
\centering \caption{Refinement results for two-phase fitting to
\ptps . ``Rietveld" and ``PDF" refer to Rietveld and PDF fits,
respectively, where the composition of the two phases was fixed to
PbTe and PbS.
   $n$ and $n_{0}$ refer to the refined and expected
(based on stoichiometry) phase fractions for the PbS-rich phase}
\begin{tabular}{|c|c|c|c|c|c|c|c|c|}
\hline  \multicolumn{1}{|c}{} & \multicolumn{2}{|c|}{$x=0.25$} &
\multicolumn{2}{c|}{$x=0.5$}& \multicolumn{2}{c|}{$x=0.75$} \\
\hline
&Rietveld&PDF&Rietveld&PDF&Rietveld&PDF\\
\hline $R_{w}$ &0.03427  &0.118 & 0.0468 &0.151418 &0.03385 & 0.0996 \\
$n/n_{0}$ & 0.19/0.25 & 0.20/0.25 & 0.50/0.50 & 0.49/0.50 &0.71/0.75 & 0.85/0.75\\
\hline C & 6.4669(3) &6.446(3) &  6.4418(3)
 & 6.414(3)  &  6.4301(3)& 6.415(3)\\
$U_{Pb}$  (\AA$^2$)&0.037(6) &0.040(4)  &0.041(6) &0.040(5)&
 0.040(7)& 0.040(5)\\
$U_{Te}$ (\AA$^2$) &0.015(6)  &0.016(4) &0.0052(6) & 0.019(4)&
 0.033(7)  & 0.02(4)\\
\hline  $a_{PbS}$(\AA) &5.9768(3) & 5.97(1)&5.9841(3)
 &5.953(4) & 5.9738(3)&5.956(3) \\
$U_{Pb}$ (\AA$^2$)& 0.044(8) &0.027(5) &0.034(7)&0.025(4) &
   0.024(6)  &0.023(3) \\
$U_{S}$ (\AA$^2$)&0.073(8) & 0.03(5) &-0.0027(7) & 0.031(4)&
0.0065(6) & 0.029(3) \\
\hline
\end{tabular}
\label{tab;PDF-Rietveld-Nominal}
\end{table*}
%%%%%%%%%%%%%%%%%%%%%%%%%%%%%%%%%%%%%%%%%%%%%%%%%%%%%%%%%%%%%%%%%%%
%  End Table 2
%%%%%%%%%%%%%%%%%%%%%%%%%%%%%%%%%%%%%%%%%%%%%%%%%%%%%%%%%%%%%%%%%%%
%
The two-phase fits of pure PbS and PbTe are good (``Rietveld" and
``PDF" columns in the table), as indicated by the low residuals. The
refined atomic displacement parameters (ADPs) are also in good
agreement with the end-member refinements, though the refinements of
this parameter are somewhat unstable on the PbS phase when it is the
minority phase as it does not contribute strongly to the scattering
in that case.  The result that relatively large ADPs are needed on
the Pb site in PbTe and on the S site in PbS are reproduced in the
two-phase fits of the phase separated samples.

The lattice parameters of the PbTe in the phase separated samples
are consistently shorter than for the pure material, and they are
consistently longer for the PbS phase component.  This effect is
real and reflects the fact that the phases in the phase separated
samples are actually solid-solutions with finite amounts of S in the
PbTe and Te in the PbS phase, respectively. We can make a rough
estimation of the composition of the phase separated phases by
considering their refined lattice parameters and assuming that
Vegard's law\cite{vegar;zp21,thorp;prb90} is obeyed in the vicinity
of the end-member compositions.  In this case, the formula for the
lattice parameter in the solid solution of composition
PbTe$_{(1-y)}$S$_y$ is $a_{y} = y(a_{\rm PbTe})+(1-y)a_{\rm PbS}$.
Thus, we can estimate the compositions of the solid-solutions in the
phase separated phases from the Rietveld refined lattice parameters.
We find that in the $x=0.25$ phases, $y=0.94$ for the PbS rich phase
and $y=0.05$ for the PbTe rich phase. This verifies that the
composition of the phases in the two-phase mixture are indeed very
near PbTe and PbS.  The values determined from the $x=0.5$ and 0.75
samples give nearly the same result with the estimated composition
of the PbTe-rich phase as $y=0.895$ and that of PbS  $y=0.03$. These
numbers are consistent with estimates from TEM evidence of a solid
solubility limit of 3\% .\cite{andro;unpub06} A more precise
determination of these values would be desirable.

 The powder diffraction data are
relatively insensitive to small changes in chemical composition of
the particular phases\cite{proff;zk02} which explains the good fit
to the data with the end-member PbS and PbTe compositions, albeit
with modified lattice constants. However, for completeness we have
carried out two-phase refinements to the phase separated data using
the nominal compositions for the two phases that were determined
above. The fits were comparable in quality to those where the
composition of the two phases were limited to pure PbTe and PbS, and
more physical ADPs were refined on the PbS component.

The agreement of the refined with the nominal composition, $n/n_0$,
is best in the $x=0.50$ sample in both the PDF and Rietveld data. It
is less good, though acceptable for the 0.25 and 0.75. Due to the
relative insensitivity to chemical composition we expect rather
large error bars on these quantities and do not ascribe significance
to the differences.  The agreement between the Rietveld and PDF
results shows that the phase separation is macroscopic since we get
the same result in both the local and average structures.

\begin{figure}[tb]
\centering
\includegraphics[width=0.45\textwidth,angle=270]{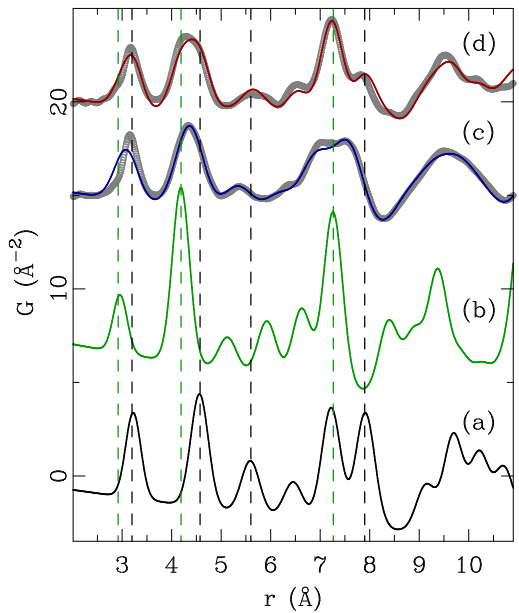}
\caption{PDFs of converged models for (a) $x=0.00$ and (b) $x=1.00$
\ptps ~samples. Comparison of the data for (c) quenched and (d)
unquenched $x=0.50$ samples (open symbols) with the solid solution
(c) and mixture (d) models (solid lines), respectively. See text for
details. Vertical dashed lines indicate positions of selected PDF
features characteristic for the end-member compositions, for easier
comparison.} \label{fig;50pctsimulation}
\end{figure}

We now consider the samples that appear from the qualitative
analysis of the data to be solid-solutions: $x=0.5$ (quenched) and
$x=0.16$.  In Fig.~\ref{fig;50pctsimulation} we consider the $x=0.5$
sample.
%
%%%%%%%%%%%%%%%%%%%%%%%%%%%%%%%%%%%%%%%%%%%%%%%%%%%%%%%%%%%%%%%%%%%
% Start Figure 7
%%%%%%%%%%%%%%%%%%%%%%%%%%%%%%%%%%%%%%%%%%%%%%%%%%%%%%%%%%%%%%%%%%%
%%%%%%%%%%%%%%%%%%%%%%%%%%%%%%%%%%%%%%%%%%%%%%%%%%%%%%%%%%%%%%%%%%%
% End Figure 7
%%%%%%%%%%%%%%%%%%%%%%%%%%%%%%%%%%%%%%%%%%%%%%%%%%%%%%%%%%%%%%%%%%%
%
In this figure, model PDFs of the undoped end-members are reproduced
for reference and the positions of their main peaks are marked.  The
quenched data are shown as grey symbols in the curves (c) and the
annealed data in the curves (d).  The magenta lines are simulated
PDFs.  In (c) the simulated PDF is from a homogeneous solid-solution
virtual-crystal model with the right nominal composition and lattice
parameter.  It agrees well with the data. In (d) the simulated PDF
is a linear combination of the PbTe and PbS PDFs.  In each case the
ADPs of the simulations have been adjusted to give the best
agreement with the data.  The simulations fit rather well indicating
that this picture of phase separation (annealed) vs solid solution
(quenched) is a good explanation of the bulk behavior for the
$x=0.5$ sample. Quantitative refinement results for the quenched
50\% sample are reproduced in Table~\ref{tab;quenched}
%
%%%%%%%%%%%%%%%%%%%%%%%%%%%%%%%%%%%%%%%%%%%%%%%%%%%%%%%%%%%%%%%%%%%
%  Start Table 3
%%%%%%%%%%%%%%%%%%%%%%%%%%%%%%%%%%%%%%%%%%%%%%%%%%%%%%%%%%%%%%%%%%%
\begin{table}[tbp]
\centering \caption{Refinement results from both PDF and Rietveld
for the quenched 50\% sample from a homogeneous solid-solution
model.}
%\nb{please fill in the missing values}
\begin{tabular}{|c|c|c|}
\hline & Rietveld & PDF  \\
\hline  $R_{w}$ & 0.047  & 0.163  \\
$a$~(\AA)& 6.2571(4)& 6.217(3)  \\
$U_{Pb}$~(\AA$^2$)& 0.055(5)& 0.062(3)\\
$U_{Te,S}$~(\AA$^2$)& 0.017(5)& 0.054(3) \\
\hline
\end{tabular}
\label{tab;quenched}
\end{table}
%%%%%%%%%%%%%%%%%%%%%%%%%%%%%%%%%%%%%%%%%%%%%%%%%%%%%%%%%%%%%%%%%%%
%  End Table 3
%%%%%%%%%%%%%%%%%%%%%%%%%%%%%%%%%%%%%%%%%%%%%%%%%%%%%%%%%%%%%%%%%%%
%
The fits are good with low $R_w$'s and reasonable refined
parameters.  The refined lattice parameter is between the end-member
values as expected and the ADP on the Pb-site is further enlarged
from the end-member values as expected due to disorder in the alloy.

\begin{figure}[tb]
\centering
\includegraphics[width=0.35\textwidth,angle=0]{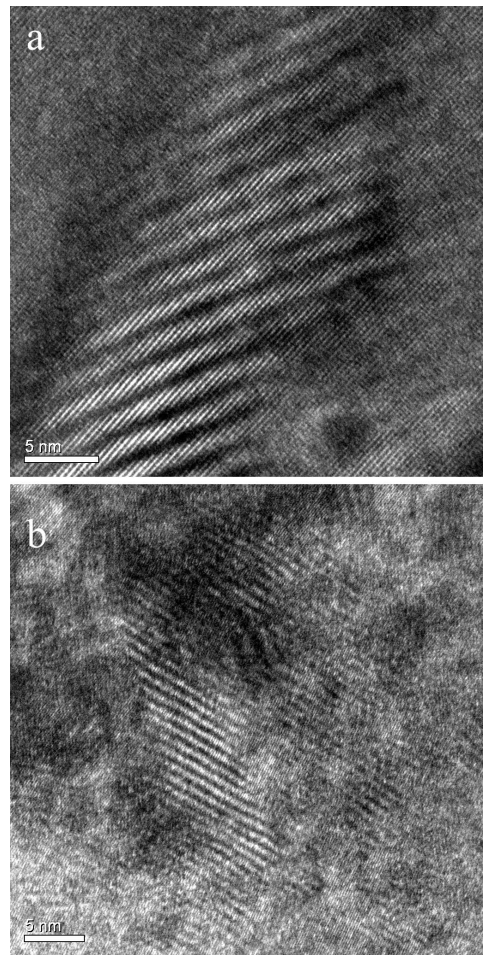}
\caption{HRTEM images of (a) $x=0.16$ and (b) quenched $x=0.50$
\ptps\ samples.} \label{fig;HRTEM}
\end{figure}

In the quenched $x=0.5$ sample the solid-solution is not
thermodynamically stable but is metastably trapped by the rapid
quench.  The quench is mostly successful at suppressing phase
separation as discussed above. However, it is not completely
successful, as TEM images of the quenched $x=0.5$ sample indicate
that the sample has compositional modulations, as shown in
Fig.~\ref{fig;HRTEM}(b).
%
%%%%%%%%%%%%%%%%%%%%%%%%%%%%%%%%%%%%%%%%%%%%%%%%%%%%%%%%%%%%%%%%%%%
% Start Figure 8
%%%%%%%%%%%%%%%%%%%%%%%%%%%%%%%%%%%%%%%%%%%%%%%%%%%%%%%%%%%%%%%%%%%
%%%%%%%%%%%%%%%%%%%%%%%%%%%%%%%%%%%%%%%%%%%%%%%%%%%%%%%%%%%%%%%%%%%
% End Figure 8
%%%%%%%%%%%%%%%%%%%%%%%%%%%%%%%%%%%%%%%%%%%%%%%%%%%%%%%%%%%%%%%%%%%
%
The striped nature of these modulations suggests that there is an
arrested spinodal decomposition taking place in the 50\% doped
sample, that would result in sinusoidal compositional modulations
about the nominal 50\% composition.  The amplitude of the
modulations is not known, but the good agreement of the homogeneous
solid-solution model to the PDF and Rietveld data suggest that the
variation in composition around the nominal 50\% is not too large.

Thus we understand the quenched 50\% sample to be close to an ideal
metastable solid solution, but with an arrested spinodal
decomposition that gives rise to nano-scale compositional
modulations.  The extent of the spinodal decomposition in the
quenched $x=0.5$ sample is difficult to assess.

Of greater interest from both a technological and scientific
viewpoint is the behavior of the $x=0.16$ sample that shows
especially good thermoelectricity. As discussed above, the
diffraction data in Fig~\ref{fig;FqAllLowQ} suggests that the sample
is macroscopically a solid solution even though it lies outside the
range of solid solubility suggested by the phase
diagrams\cite{darro;tmsaime66,leute;zpc85} and inferred from the
composition of the PbTe-rich phase of the phase-separated
compositions in our own refinements (25\%, 50\%, 75\% sample).

\begin{figure}[tb]
\centering
\includegraphics[width=0.35\textwidth,angle=270]{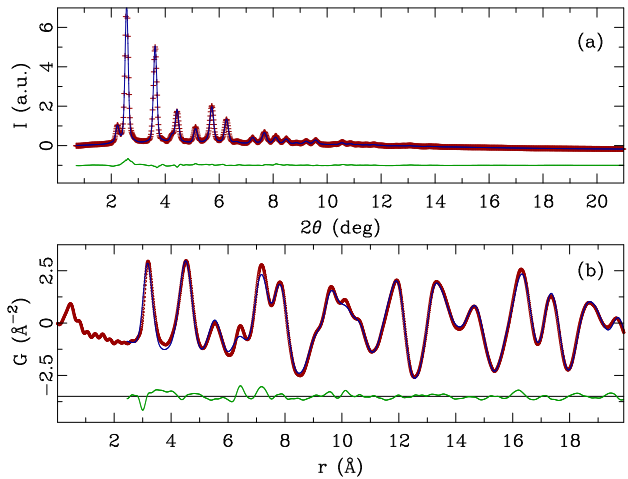}
\caption{Representative refinements of the $x=0.16$ sample data
using (a) Rietveld and (b) PDF approach. Symbols represent data,
and solid lines are the model fits. The difference curves are offset
for clarity.}
\label{fig;PbTe84-PbS16}
\end{figure}

We tried fitting two-phase and homogeneous models to both the
diffraction and PDF data. The results are shown in
Table~\ref{tab;Compare-Threemodels} with representative fits shown
in Fig.~\ref{fig;PbTe84-PbS16}.
%
%%%%%%%%%%%%%%%%%%%%%%%%%%%%%%%%%%%%%%%%%%%%%%%%%%%%%%%%%%%%%%%%%%%
% Start Figure 9
%%%%%%%%%%%%%%%%%%%%%%%%%%%%%%%%%%%%%%%%%%%%%%%%%%%%%%%%%%%%%%%%%%%
%%%%%%%%%%%%%%%%%%%%%%%%%%%%%%%%%%%%%%%%%%%%%%%%%%%%%%%%%%%%%%%%%%%
% End Figure 9
%%%%%%%%%%%%%%%%%%%%%%%%%%%%%%%%%%%%%%%%%%%%%%%%%%%%%%%%%%%%%%%%%%%
%
%%%%%%%%%%%%%%%%%%%%%%%%%%%%%%%%%%%%%%%%%%%%%%%%%%%%%%%%%%%%%%%%%%%
%  Start Table 4
%%%%%%%%%%%%%%%%%%%%%%%%%%%%%%%%%%%%%%%%%%%%%%%%%%%%%%%%%%%%%%%%%%%
\begin{table*}[tbp]
\centering \caption{Rietveld and PDF refinement results from three
different models for the \ptpstwo\ sample: model A is solid solution
model, model B is a simple two-phase mixture of PbTe and PbS and
model C is a mixture of pure PbTe phase plus a solid solution of
composition PbTe$_{0.5}$-PbS$_{0.5}$. $n$ and $n_{0}$ refer to the
refined and expected (based on stoichiometry) phase fractions for
the PbS-rich phase.}
\begin{tabular}{|c|c|c|c|c|c|c|c|}
\hline  \multicolumn{2}{|c}{} & \multicolumn{2}{|c|}{model A} &
\multicolumn{2}{c|}{model B} &
\multicolumn{2}{c|}{model C} \\
\hline  \multicolumn{2}{|c|}{} & Rietveld & PDF & Rietveld & PDF & Rietveld & PDF \\
\hline  & $R_{w}$ & 0.04647  & 0.1209& 0.05186 & 0.121& 0.03068 &0.114\\
 & $n/n_{0}$ & -- & --& 0.14/0.16 & 0.037/0.16 & 0.31/0.32 & 0.24/0.32 \\
\hline PbTe & $a$ (\AA) & 6.4264(5) & 6.403(3)& 6.4233(4)
& 6.403(24)&6.4203(4)&6.416(3) \\
  & $U_{Pb}$ (\AA$^2$)& 0.047(5) & 0.047(3) & 0.035(6) & 0.035(3)&
 0.028(6)&0.036(4)\\
  & $U_{Te}$ (\AA$^2$)&  0.0061(6) & 0.019(3)& 0.023(6) & 0.029(4)&
  0.016(6) & 0.025(5) \\
\hline second phase & $a$ (\AA) & -- & -- & 5.900(1) & 5.942(4)&
6.1673(3)&6.255(3) \\
 & $U_{Pb}$  (\AA$^2$)& -- & --& 0.018(8) & 0.021(6)&  0.253(8)&0.064(6)  \\
  & $U_{S,Te}$  (\AA$^2$)& -- &  -- & 0.013(8) &
  -0.0024(6)&0.253(8) & 0.070(6)\\
\hline
\end{tabular}
\label{tab;Compare-Threemodels}
\end{table*}
%%%%%%%%%%%%%%%%%%%%%%%%%%%%%%%%%%%%%%%%%%%%%%%%%%%%%%%%%%%%%%%%%%%
%  End Table 4
%%%%%%%%%%%%%%%%%%%%%%%%%%%%%%%%%%%%%%%%%%%%%%%%%%%%%%%%%%%%%%%%%%%
%
As expected from the qualitative analysis of the data discussed
above, the single-phase solid-solution model (model A) provides
acceptable fits to the data.  The refined lattice parameters are
shorter than pure PbTe.  According to the Vegard's law analysis, the
refined lattice parameter gives a nominal composition for this
sample of 0.14 (Rietveld)/0.12(PDF), in reasonable agreement with
the actual composition.   Enlarged ADPs are found on the Pb
sublattice with smaller ADPs on the Te lattice, as was the case for
the PbTe end-member. As expected for a solid-solution, the ADPs are
enlarged with respect to PbTe.

For completeness, we also tried the simple model of phase separation
into pure PbTe and PbS end-members.  The results appear in
Table~\ref{tab;Compare-Threemodels} as model B.  The Rietveld fit is
significantly worse as measured by $R_w$.  In the case of the PDF
fit the $R_w$ is comparable but the refinement reduced the phase
fraction of the second phase and adjusted the lattice parameter of
the majority phase, moving the refinement back towards the
solid-solution result. This refinement also returned unphysical
negative atomic displacement factors on the minority phase.  The
solid-solution model is clearly preferred over full phase separation
from the bulk diffraction measurements.

The TEM images from the 16\% sample (Ref.~\onlinecite{andro;unpub06}
and Fig.~\ref{fig;HRTEM}(a)) suggest that it is two-phased, with one
phase being homogeneous and the other resembling the quenched
$x=0.5$ sample with arrested spinodal decomposition.  A model that
simulated this situation was successful compared to the PDF data, as
shown from model C in Table~\ref{tab;Compare-Threemodels}.  This
model assumed that the nominally 16\% sample is phase separated into
regions that are pure PbTe and regions that resemble the quenched
50\% sample, i.e., they are nominally $x=0.5$ solid-solutions but
also exhibiting spinodal decomposition as suggested by the TEM
images. Thus, model C is a phase separation into pure PbTe and a
solid solution of composition PbTe$_{0.5}$S$_{0.5}$. This model
gives the lowest $R_w$'s for fits to the 16\% compound in both the
Rietveld and PDF refinements.  The phase fractions were free to vary
but refined to values that are close to those expected. The lattice
constants refined to reasonable values. The majority phase lattice
constant was close to that of the PbTe rich phase in the two-phase
refinements in Table~\ref{tab;PDF-Rietveld-Nominal}.  In the case of
the minority phase, the lattice constant lay between pure PbTe and
PbS consistent with a nominal 50\% composition.  The ADPs are
slightly large in the PbTe-rich phase but physically reasonable. In
the minority phase the ADPs are unphysical in the Rietveld
refinement suggesting that this parameter is not well determined in
the refinement. However, in the PDF refinement they are more
reasonable, but very large.  This is perfectly consistent with the
fact that this minority phase itself actually has a compositional
variation due to the spinodal effects.

\section{Summary}
This work confirmed the phase separation tendency of the PbTe/PbS
system.  It also showed that phase separation can be effectively,
but not completely, suppressed by quenching at 50\% composition,
where a partial spinodal decomposition appears to be taking place,
at least in a portion of the sample.

However, the main result is an improvement in our understanding of
the state of the thermoelectrically promising 16\% sample.
Measurements of the bulk average structure, and the bulk local
structure, indicate that it is not phase separated into PbTe-rich
and PbS-poor end-members like the other similarly processed samples
in the series. The best explanation of all the data at hand is that
this sample prefers a phase separation into a PbTe-rich phase and a
phase that is nominally 50\% doped, but which has a partial spinodal
decomposition reminiscent of the quenched 50\% sample. Such a
nano-scale phase separation is thought to be important in producing
the very low lattice thermal conductivity $\kappa$ that is observed
in this material\cite{andro;unpub06}.  Interestingly, in this case
the effect appeared not after a quench, but after an anneal,
suggesting that it is the thermodynamically preferred state, though
this needs to be investigated further. Also of interest is to
explore further the nature of the PbTe-rich component, which
preliminary TEM investigations\cite{andro;unpub06} indicate also
contains nanostructured regions with nano-scale nuclei of a second
phase present.

The other important observation from this work is that quenching is
very important in determining the phase separation and resulting
nano-scale microstructure.  This suggests that in this system it may
be possible to engineer $\kappa$, and therefore $ZT$ in the bulk
material by appropriate heat treatments.  This is a promising route
for future research.

%
%%%%%%%%%%%%%%%%%%%%%%%%%%%%%%%%%%%%%%%%%%%%%%%%%%%%%%%%%%%%%
% Acknowledgements
%%%%%%%%%%%%%%%%%%%%%%%%%%%%%%%%%%%%%%%%%%%%%%%%%%%%%%%%%%%%%
%
\acknowledgements{We are grateful to Douglas Robinson for kind help
with the experimental setup. We acknowledge Pavol Juhas, Ahmad
Masadeh, HyunJeong Kim, and Asel Sartbaeva for their valuable
assistance with the data collection. This work was supported in part
by National Science Foundation (NSF) grant DMR-0304391 and by the
Office of Naval Research. Data were collected at the 6IDD beamline
in the Midwest Universities Collaborative Access Team (MUCAT) sector
at the Advanced Photon Source (APS).  Use of the Advanced Photon
Source was supported by the U. S. Department of Energy, Office of
Science, Office of Basic Energy Sciences, under Contract No.
DE-AC02-06CH11357. The MUCAT sector at the APS is supported by the
U.S. DOE, Office of Science, Office of Basic Energy Sciences,
through the Ames Laboratory under Contract No. W-7405-Eng-82.}
%
%%%%%%%%%%%%%%%%%%%%%%%%%%%%%%%%%%%%%%%%%%%%%%%%%%%%%%%%%%%%%
%End Acknowledgements
%%%%%%%%%%%%%%%%%%%%%%%%%%%%%%%%%%%%%%%%%%%%%%%%%%%%%%%%%%%%%
%

%
%\bibliography{billinge-group,%

\begin{thebibliography}{35}
\bibitem[{ZTn()}]{ZTnote}
\bibinfo{note}{$ZT= \sigma S^{2} T /\kappa $, where $T$ is the operating
  temperature of the device, $S$ is the thermopower, and $\sigma$ and $\kappa$
  are the electrical and thermal conductivities of the material, respectively}.
\end{thebibliography}
%             abb-billinge-group,%
%              everyone%
%}
%\bibliographystyle{apsrev}

\end{document}